# Localized vs. delocalized character of charge carriers in $LaAlO_3$/ $SrTiO_3$ superlattices

Kejin Zhou[1], Milan Radovic[2,1], Justine Schlappa[1], Vladimir Strocov[1], Ruggero Frison[3], Joel Mesot[1,2], Luc Patthey[1], and Thorsten Schmitt[1]*

[1]Paul Scherrer Institut, Swiss Light Source, CH-5232 Villigen PSI, Switzerland
[2]Laboratory for synchrotron and neutron spectroscopy, Ecole Polytechnique Federale de Lausanne, CH-1015 Lausanne, Switzerland
[3]Laboratory for Developments and Methods, Paul Scherrer Institut, CH-5232 Villigen PSI, Switzerland

* E-mail: thorsten.schmitt@psi.ch

## Abstract

Understanding the nature of electrical conductivity, superconductivity and magnetism between layers of oxides is of immense importance for the design of electronic devices employing oxide heterostructures. We demonstrate that resonant inelastic X-ray scattering can be applied to directly probe the carriers in oxide heterostructures. Our investigation on epitaxially grown $LaAlO_3/SrTiO_3$ superlattices unambiguously reveals the presence of both localized and delocalized Ti 3d carriers. These two types of carriers are caused by oxygen vacancies and electron transfer due to the polar discontinuity at the interface. This result allows explaining the reported discrepancy between theoretically calculated and experimentally measured carrier density values in $LaAlO_3/SrTiO_3$ heterostructures.

Atomically flat artificial heterostructures from complex perovskite oxides display exotic phenomena not existing in the bulk of each of the constituents (*1-3*). Even a relatively simple interface between the band insulators $LaAlO_3$ (LAO) and $SrTiO_3$ (STO), exhibits a rich phase diagram. Building polar LAO upon non-polar $TiO_2$-terminated STO leads to metallicity, magnetism and two-dimensional superconductivity at the n-type interfaces (*4-8*). These unexpected properties spurred numerous investigations on the origin of this interface conductivity. Electronic reconstruction, i.e. partial change of Ti ions from 4+ to 3+ oxidation state by electron transfer from LAO, may occur in this kind of oxide heterostructures. Through this process the polar discontinuity at the interface is solved (*9-10*). However, the real mechanism is more complex than initially suggested in this ionic picture. A change of carrier density by about three orders of magnitude upon the variation of oxygen deposition pressure ($P_{O2}$) highlights the importance of oxygen vacancies (OV) in the creation of the interface conductivity (*11-13*). Indeed, it was found by transport measurements that doping of the LAO/STO with OV created during the growth induces metallic behavior (*12*). On the other hand, appearance of conductivity above a critical thickness of LAO indicates that the electron transfer scenario should contribute as well (*14*). If the electron transfer is the major mechanism (half an electron transferred per unit cell) for system stabilization, the sheet carrier density would have to be $3.3 \times 10^{14}$ cm$^{-2}$. This is in strong disagreement with the observed experimental value which is nearly one order of magnitude lower. To reconcile this one can propose that not necessarily all transferred electrons contribute to conductivity. Indeed, Density Functional Theory (DFT) calculations proposed that the electrons occupy localized as well as delocalized Ti subbands and only the latter contribute to transport (*15*). Other DFT calculations suggest that, apart from the electronic reconstruction, the polar discontinuity can be partially compensated by the structural distortion (*16*). Moreover, surface X-ray diffraction revealed a significant role of cationic inter-mixing at the interface (*17*). Those studies suggest that during the growth and post-annealing of LAO/STO heterostructures a complicated mechanism consisting of several processes may be at play. Unravelling the electronic structure of buried oxide-interfaces in such heterostructures is, therefore, very important to clarify the possible mechanism and to determine the main driving force for creating the metallic interface.

In fact, several spectroscopic techniques have recently been used to determine the electronic structure of LAO/STO heterostructures. Resonant soft X-ray scattering found spectroscopic evidence for interface electronic reconstruction of Ti 3d and O 2p states (*18*). Electron energy loss spectroscopy detected sizable amount of $Ti^{3+}$ states at n-type interfaces (*10*). Ti 2p-edge X-ray absorption spectroscopy (XAS) was employed to characterize the orbital reconstruction of $Ti^{4+}$ sites, although no localized $Ti^{3+}$ carriers were discovered (*19*). In contrast, localized $Ti^{3+}$ carriers were observed with core level hard X-ray photoelectron spectroscopy (*20*), however, soft X-ray photoemission investigations did not reveal any indication for occupied Ti 3d states (*21*). In order to understand the metallic origin of the LAO/STO interface, it is therefore essential to use a probe with particular sensitivity to valence Ti 3d states.

In this report, we demonstrate how resonant inelastic X-ray scattering (RIXS) at the Ti $2p_{3/2}$ edge gives direct information on the electronic structure of the occupied Ti 3d states in LAO/STO heterostructures. RIXS is as a non-destructive and volume sensitive photon-in/photon-out technique (attenuation length ~ 80 nm at Ti 2p edge (*22*)) ideal for investigating buried interfaces (*23*). In our investigation we make use of the fact that the RIXS process at the Ti $2p_{3/2}$ edge in the presence of an empty 3d shell, i.e. $Ti^{4+}$ ($d^0$ configuration) cations, only gives rise to an elastic emission signal, when the excited $2p_{3/2}$ core electrons recombine with the core hole. On the other hand, for $Ti^{3+}$ ($d^1$) cations the single 3d electron opens additional decay channels giving rise to inelastic emission (dd excitations) (see the scheme in Fig. 1A) (*24*). The intensity of inelastic emission at the Ti $2p_{3/2}$ edge is a direct measure for the amount of Ti 3d electrons despite the majority of $Ti^{4+}$ cations in the LAO/STO system. Taking advantage of this, we can understand whether 3d charge carriers exist at all and how they are coupled to the origin of the interfacial metallicity.

The RIXS experiments (see setup in Fig. 1B) were performed at the ADRESS beam-line (*25*) of the Swiss Light Source at the Paul Scherrer Institut on a series of LAO/STO superlattices (SLs) grown with a Pulsed Laser Deposition (PLD) set-up at the SIS beam-line. All samples contain 10 LAO/STO stacking periods, in which the thickness of LAO layers is increased from 3 to 6, 8 and 10 unit cells (uc), while STO layers are fixed to 10 uc. In the following, SL samples are denoted as LAO*m*, where *m* stands for the number of LAO uc in each layer. Growth and annealing procedures are described in (*26*).

Our analysis begins with a LAO*3* SL which contains a thickness of LAO layers below the observed threshold for the insulator-metal transition in bilayers (*14*). Fig. 2A displays a total fluorescence yield Ti 2p XAS spectrum from the LAO*3* SL. For different incident energies $\Omega$ in the range of the $Ti^{3+}$ $2p_{3/2}$-$e_g$ threshold (*19*), RIXS spectra were measured and depicted in Fig. 2B. Besides the elastic peaks at zero energy position, low energy excitations around an energy loss of 300 meV are seen as a shoulder in all spectra. Furthermore, a double peak excitation shows up around 2.8 eV. In the RIXS process, excitations at constant energy loss result from local excitations like crystal field excitations (dd-excitations) (*23*). The 2.8 eV peak is known from $LaTiO_3$ single crystals with titanium atoms in the $Ti^{3+}$ state, to be caused by localized dd excitations between the on-site $t_{2g}$ and $e_g$ orbitals. The 300 meV peak stems from intra-$t_{2g}$ excitations (*24*).

As described above, dd excitations are a clear signature for localized Ti 3d electrons and we can therefore conclude that such electrons are created in the LAO*3* SL. Very interestingly, a remarkable dispersive excitation ~ 1-2 eV is observed in all spectra not showing a constant energy loss as localized ones. The non-constant energy loss of this excitation is clear evidence for scattering from delocalized electrons (*23*). RIXS spectra from other as-grown SLs with increasing thickness of LAO layers show the same localized and delocalized excitations as displayed by LAO*3*. For a reference we performed RIXS on a pure $TiO_2$-terminated STO substrate treated in the same way as the SLs. This reference sample just delivers a negligible background signal, which indicates that Ti 3d carriers must be generated during the SL growth. To exclude that carriers are created in STO films during the growth, RIXS was measured on a 100 uc thick homoepitaxial STO film (equal STO volume as in the SLs) grown under exactly the same conditions as the SLs. However, the STO film provides just delocalized excitations with significantly lower intensity compared to LAO*3*. From the difference between STO substrate and film on the one side and LAO3 on the other, we conclude that the occurrence of both types of excitations in SLs is induced by the building of LAO upon STO. This is in strong contrast to RIXS spectra of bulk $LaTiO_3$ which are only characterized by localized dd excitations (*24*). We thus conclude that in LAO/STO heterostructures, two types of electrons are created: ones bound to $Ti^{4+}$ forming localized $Ti^{3+}$ cations, and others in the vicinity of the $Ti^{4+}$ cation producing unpaired delocalized electrons. Our results give experimental proof for the proposal by Popović *et al.* (*15*) and a discussion by other studies on LAO/STO heterostructures (*6, 27-30*) that localized and delocalized types of carriers occupy different Ti bands with the latter having higher mobility.

The creation of two types of Ti 3d electrons depends strongly on the thickness of LAO layers (*m*). This is experimentally supported by comparing RIXS spectra excited at incident energy of 459.2 eV from all as-grown SLs (Fig. 2C). It is clear that the intensities of both localized and delocalized excitations increase considerably with the LAO thickness, while their energy positions are unchanged. This proves that the crystal field splitting of $Ti^{3+}O_6$ octahedra does not change with increasing number of LAO uc. Additional LAO uc simply add Ti 3d electrons preserving the local crystal field environment. However, the origin of these Ti 3d electrons is unclear, since both OV and transferred electrons due to the polar discontinuity can be donors.

To clarify the origin of the observed carriers, all samples were annealed in oxygen in order to remove OV. RIXS spectra from annealed LAO*6* and LAO*10* SLs are displayed in Fig. 2C. Notably, the annealing reduces significantly the intensities of both types of excitations owing to the reduction of $Ti^{3+}$ to $Ti^{4+}$, while the energy positions are not altered. To gain a quantitative understanding, the localized and delocalized excitations in the spectra of all as-grown and annealed samples were deconvoluted and integrated, respectively. The resultant spectral weights shown in Fig. 3 are proportional to the carrier density because of unchanged local crystal structure among all SLs before and after annealing. The spectral weight of both localized and delocalized electrons of as-grown SLs increases monotonically with the thickness of LAO layers. After annealing, it is significantly reduced, which is in-line with the general trend of the reduction of carrier density upon oxygen annealing (*11,20*). The RIXS measurement on the reference STO film after annealing shows no change of delocalized excitations indicating that carriers inside of STO originate from intrinsic defects generated during the growth (*31-32*). Based on these observations we conclude that the Ti 3d carriers in as-grown SLs are mostly caused by OV which are created during the growth of LAO layers and then migrate into the LAO/STO interface region. Although lower oxygen migration energy of STO compared to LAO may be one of the driving forces for this effect (*11*), it might be even more boosted since LAO is strongly tensile strained upon STO (*33*).

The RIXS spectral weight after annealing provides further insight into the origin of Ti 3d electrons. From Fig. 3 one can notice that upon annealing the delocalized spectral weight of LAO*3* and LAO*6* is reduced to the same level as in the annealed STO film. A larger localized spectral weight is, however, present for LAO*3* and LAO*6*. Above LAO*6*, both localized and delocalized spectral weights are increasing and reach a plateau. Similar trends also exist in RIXS spectra recorded at other incident energies. This resembles the behavior of the electrical conductivity which shows a sharp rise beyond a critical thickness of LAO as explained by the electron transfer scenario (*14*). Since OVs are refilled after annealing, electrons transferred from LAO should be most likely responsible for the sudden increase of carrier density. Our observation of the critical thickness is in good agreement with a DFT calculation predicting that 8 uc of LAO are needed to avoid polarity catastrophe in LAO/STO SLs (*34*). Based on this we propose that during the growth of SLs the presence of OV plays a major role in the creation of Ti 3d carriers, while after the annealing process electronic reconstruction occurs in samples exceeding the critical thickness.

In summary, we demonstrated the high sensitivity of RIXS for probing $Ti^{3+}$ carriers in a large background of $Ti^{4+}$ sites at buried interfaces of LAO/STO heterostructures. RIXS distinguishes localized from delocalized Ti 3d carriers owing to their distinct energy dependent spectral response. In the same time it gives a measure of the density of localized and delocalized carriers, respectively. Our results, thereby, allow explaining the discrepancy between theoretically calculated and experimentally measured carrier density values. We ascribe the basic mechanism driving the generation of the interface conductivity to the competition of two processes, the presence of oxygen vacancies and the electron transfer due to the polar discontinuity.

**Acknowledgements**

This work was performed at the ADRESS beamline of the Swiss Light Source using the SAXES instrument jointly built by Paul Scherrer Institut, Switzerland and Politecnico di Milano, Italy. The authors acknowledge H. Y. Hwang, F. Miletto Granozio, C. Quitmann, M. Shi, and A. Kotani for stimulating discussions. L. Braicovich, G. Ghiringhelli, C. Dallera and M. Grioni are acknowledged for discussions and the collaboration within the SAXES project.

**Figure legends**

Fig. 1. (**A**) Sketch of the Ti 2p-RIXS process. $\Omega$ and $\omega$ are incident and emitted X-ray energies, respectively. The left panel describes the RIXS process for $Ti^{4+}$ states, in which $\omega=\Omega$ and only elastic emission is observed. The right panel describes RIXS for $Ti^{3+}$ states, where an extra emission channel is opened ($\omega\neq\Omega$) in addition to the elastic peak. (**B**) A schematic view of the RIXS experimental setup. The incident X-ray beam ($hv_{in}$) is linearly polarized ($\pi$ means in-plane polarization) and impinges on the sample surface with a grazing angle of 20°. The emitted X-rays ($hv_{out}$) are detected at an angle of 90° with respect to the incoming beam. Samples were aligned with the surface normal (001) in the scattering plane.

Fig. 2 (**A**) A total fluorescence yield (TFY) mode Ti L-XAS spectrum of the LAO*3* SL sample. The first two peaks are the $Ti^{4+}$ $2p_{3/2}$-$t_{2g}$ and $2p_{3/2}$-$e_g$ resonance, respectively (*19*). Arrows label the incident energies used for RIXS. (**B**) Series of RIXS spectra of LAO*3* SL excited across the $Ti^{3+}$ $2p_{3/2}$-$e_g$ threshold. RIXS spectra from STO substrate (solid black) and STO film (dashed grey) references are also displayed. (**C**) RIXS spectra after subtraction of the elastic peak from LAO*6* and LAO*10* (before and after annealing) excited at $\Omega$=459.2 eV. Dots and solid lines (thick and thin ones are for as-grown and annealed cases) stand for experimental and fitting results, respectively.

Fig. 3 (**A**) Integrated spectral weight of the excitations measured using RIXS at $\Omega$=459.2 eV. Localized and delocalized spectral weights are represented by open and filled symbols, respectively. Light blue squares and pink circles correspond to as-grown and annealed samples. Error bars represent the first standard deviation of fitting. Shaded blue and pink lines are a guide to the eye.

**Fig. 1**

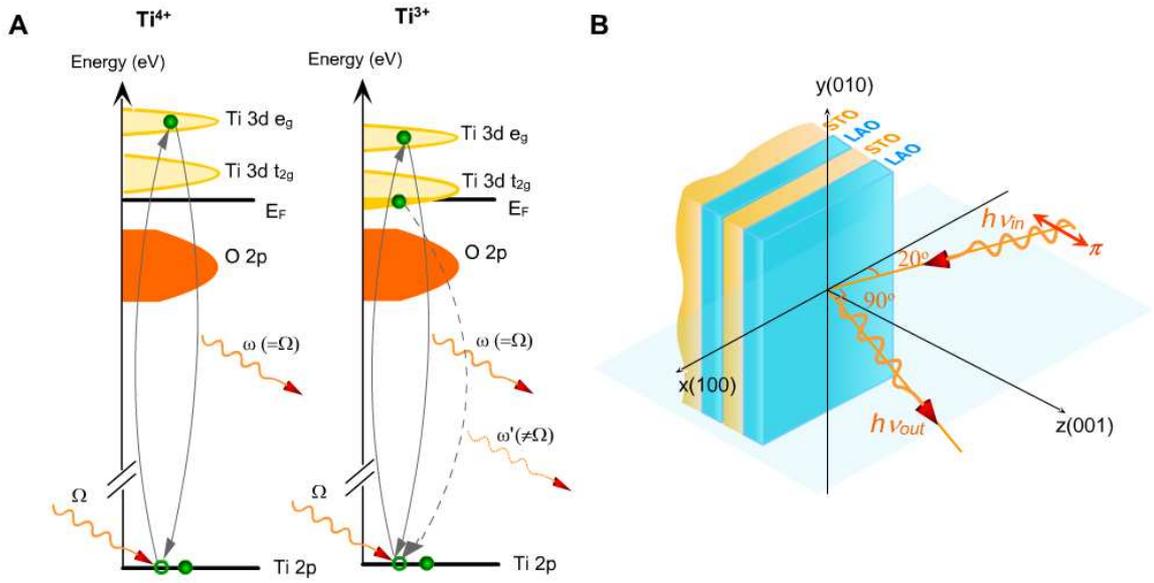

**Fig. 2**

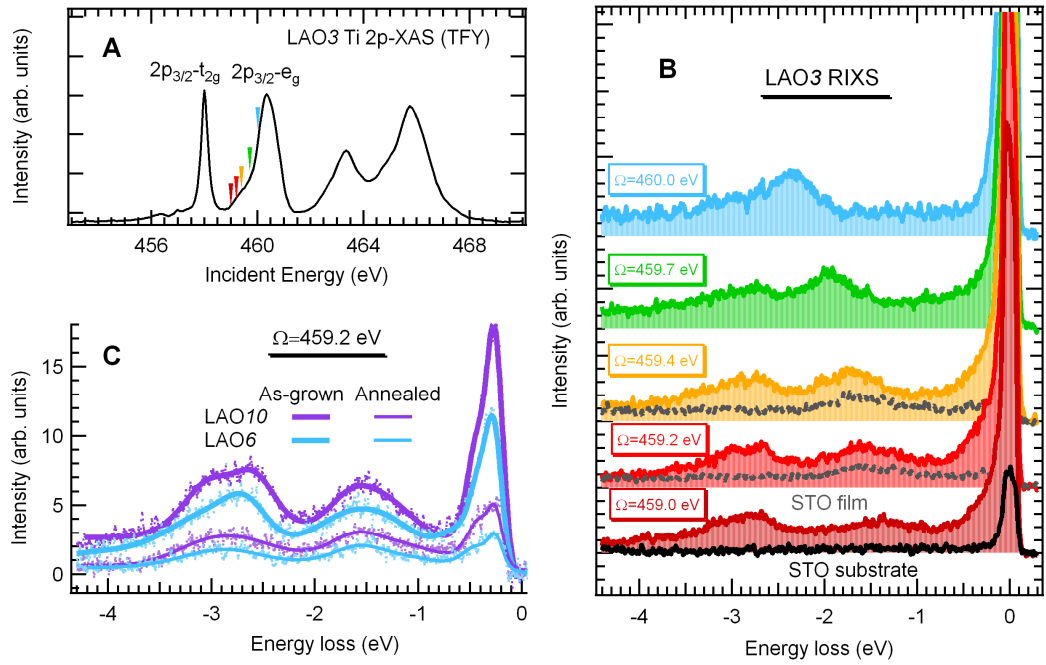

**Fig. 3**

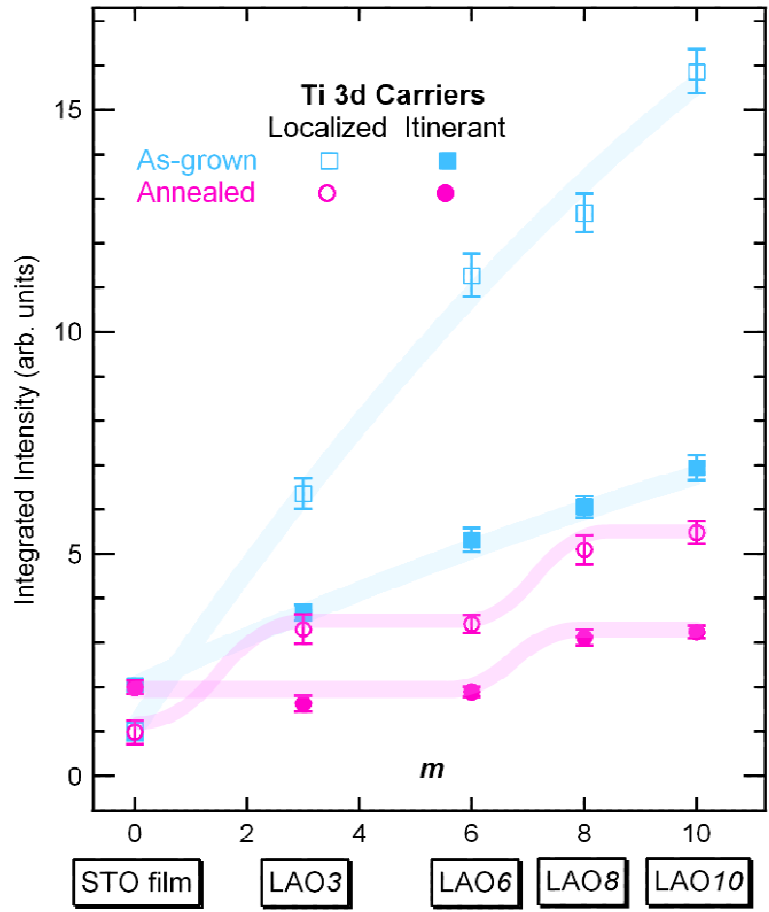